\documentclass[letterpaper,12pt]{article}

\usepackage{amssymb,latexsym,amsmath}

\usepackage{fullpage}
\usepackage{nicefrac}
\usepackage{mathrsfs}
\usepackage{slashed}
\usepackage{bbold}

\usepackage[pdftex]{graphicx}

\makeatletter \@addtoreset{equation}{section}
\makeatother

\begin{document}

\begin{titlepage}
	\thispagestyle{empty}
	\begin{flushright}
		\hfill{DFPD-12/TH/19}\\
		\hfill{ROM2F/2012/11}
	\end{flushright}
	
	\vspace{35pt}
	
	\begin{center}
	    { \LARGE{\bf de Sitter vacua in $N=8$ supergravity \\[3mm] and slow-roll conditions}}
		
		\vspace{50pt}
		
		{G.~Dall'Agata$^{1,2}$ and G.~Inverso$^{3,4}$}
		
		\vspace{25pt}
		
		{
		$^1${\it  Dipartimento di Fisica ``Galileo Galilei''\\
		Universit\`a di Padova, Via Marzolo 8, 35131 Padova, Italy}
		
		\vspace{15pt}
		
	    $^2${\it   INFN, Sezione di Padova \\
		Via Marzolo 8, 35131 Padova, Italy}
		
		\vspace{15pt}
		
		$^3${\it Dipartimento di Fisica, Universit\`a di Roma ``Tor Vergata''\\
		Via della Ricerca Scientifica, 00133 Roma, Italy}
		
		\vspace{15pt}
		
		$^4${\it  INFN, Sezione di Roma 2, ``Tor Vergata''\\
		Via della Ricerca Scientifica, 00133 Roma, Italy}
		}		
		
		\vspace{40pt}
		
		{ABSTRACT}
	\end{center}
	
	\vspace{10pt}

In this letter we discuss de Sitter vacua in maximal gauged supergravity in 4 dimensions.
We show that, using the newly deformed theories introduced in \cite{Dall'Agata:2012bb}, we can obtain de Sitter vacua with arbitrarily flat tachyonic directions in the SO(4,4)$_c$ models.

\end{titlepage}

\baselineskip 6 mm



\section{Introduction} 
\label{sec:introduction}

Astrophysical data convincingly show that our Universe went through an inflationary phase in its early epoch and is sitting today at a vacuum state with a very small, positive, vacuum energy density.
An obvious question is whether this fact is compatible with our current understanding of string theory.
In particular, we would like to understand whether vacua with a positive cosmological constant can be obtained in supergravity and string theory and what are their main features.
Focusing on supergravity, while it is rather easy to construct models with the desired properties starting from theories with minimal supersymmetry, the situation in extended theories looks grim.
Currently there are extremely few instances of metastable vacua in $N=2$ models \cite{Fre:2002pd,Roest:2009tt} and for $N=8$ we do not even have a single example where the instabilities could be used to host slow-rolling fields \cite{Kallosh:2001gr}.

In this letter we would like to address this problem in the context of maximal supergravity in 4 dimensions.
Although this theory may not lead to realistic phenomenological scenarios, its very constrained structure may help us understand why in supersymmetric theories of gravity it is so difficult to obtain positive energy vacua that are stable or, at least, where unstable directions satisfy the slow-roll conditions.

A scan of most of the known gaugings leading to de Sitter vacua \cite{Kallosh:2001gr} has led to the common belief that maximal supergravity only allows positive energy vacua with tachyons whose mass is of the order of the cosmological constant.
However, a more sophisticated analysis cannot yet exclude the possibility of having stable vacua, or at least with a better behaviour of the unstable directions \cite{Borghese:2011en}.
For this reason, we decided to revisit this question, also using some new recent developments that showed that for a given gauge group we can actually obtain an infinite number of new interesting models if we allow for non-standard embeddings of the gauge group in the duality group \cite{Dall'Agata:2012bb}.
We recall that two ingredients determine the gaugings: the choice of generators of the duality group that become local and the choice of symplectic frame, namely the choice of gauge fields that play the role of electric and of magnetic gauge fields.
In particular, it has been shown in \cite{Dall'Agata:2012bb} that the change of the parameter fixing the latter may not only change the value of the cosmological constant of the vacua found with the standard embedding, but also their very existence.
In fact for the SO(8)$_c$ models of \cite{Dall'Agata:2012bb} (as well as for SO(7,1) gaugings \cite{Borghese:2012qm}) one finds that new $SO(7)$ and $G_2$ invariant vacua appear for non-zero value of the deformation parameter $c$, also with all supersymmetries broken \cite{Dall'Agata:2012bb,Borghese:2012qm}.

As a first step, we decided to focus on the SO(4,4) gauged supergravity model, which is known to admit an unstable de Sitter vacuum, sitting at the origin of the moduli space in the standard embedding.
By performing a consistent truncation, first to the SO(3) $\times$ SO(3) singlets and then projecting further with respect to a ${\mathbb Z}_2$, we analyze a consistent truncation to a potential with only 2 scalar fields for which we discuss the vacua and the mass spectrum fully analytically.
We will see that this analysis leads to the observation of 3 de Sitter vacua in the generic $c$-deformed model.
For 2 of them both the cosmological constant and the value of the masses depend on the deformation parameter, differently from what happens in the models analyzed so far \cite{Dall'Agata:2012bb,Borghese:2012qm,Kodama:2012hu}.
In particular, we will show that, although all these vacua are always unstable, the new vacua for which the masses depend on $c$ allow for arbitrarily flat unstable directions, hence providing the first example of slow-roll conditions within maximal supergravity, i.e.~arbitrarily small ratios between the scale of the tachyonic masses and of the vacuum energy.



\section{SO(4,4)$_c$ gauged supergravity} 
\label{sec:so_4_4__c_gauged_supergravity}

It has been known for a long time that the SO(4,4) gauging of maximal supergravity allows for an unstable de Sitter vacuum \cite{HullWarner}.
In a more modern language, we can obtain this gauging in terms of the embedding tensor $\Theta_M{}^\alpha$, specifying the couplings of the electric and magnetic vector fields $A_\mu^M$, $M=1,\ldots,56$, to the E$_{7(7)}$ generators $t_{\alpha}$, $\alpha = 1,\ldots,133$, for instance in the covariant derivatives $D_\mu = \partial_\mu - A_\mu^M \, \Theta_M{}^\alpha t_{\alpha}$ \cite{de Wit:2007mt}.

In the standard SL(8,${\mathbb R}$) symplectic frame, all CSO($p,q,r$) groups (with $p+q+r=8$) can be obtained by embedding them in the SL(8,${\mathbb R}$) subgroup of E$_{7(7)}$ \cite{Cordaro:1998tx,de Wit:2007mt}.
In such a frame, the electric vector fields transform in the ${\bf 28}$ of $\mathrm{SL}(8,\mathbb{R})$, while the magnetic ones transform in the ${\bf 28}'$: $A_\mu^M = \{A_\mu^{[AB]}, A_{\mu\,[AB]}\}$, where $A,B=1,\ldots,8$ are indices labelling the fundamental representation of ${\mathfrak sl}$(8,${\mathbb R}$). 
Also the 133 generators of the E$_{7(7)}$ group in the SL$(8,{\mathbb R})$ basis can be divided according to the decomposition $\mathbf{133} \to \mathbf{63} + \mathbf{70},$ where the first 63 are the generators of the SL($8,{\mathbb R}$) subgroup of E$_{7(7)}$, which we name $t_A{}^B$, and the remaining 70 are described by a rank 4 totally antisymmetric tensor $t^{ABCD}$.
The standard SO(4,4) gauging of \cite{HullWarner} can then be reproduced by choosing the embedding tensor as 
\begin{equation}
	\label{thetacoupling}
	\Theta_{M}{}^\alpha = \Theta_{AB}{}^C{}_D \propto \delta_{[A}^C \theta^{\vphantom{C}}_{B]D},
\end{equation}
where $\theta_{AB}$, which couples the electric vectors to the SL(8,${\mathbb R}$) generators $t_C{}^D$, is chosen to be a diagonal metric of the form \cite{DallAgata:2011aa,Kodama:2012hu}
\begin{equation}
	\label{theta}
	\theta = {\rm diag} \{1,1,1,1,-1,-1,-1,-1\}.  
\end{equation}
However, as noted in \cite{Dall'Agata:2012bb}, the generic decomposition of the representation $\mathbf{912}$ of E$_{7(7)}$ with respect to SO(8) contains two singlets that can be used as invariant tensors describing the corresponding gauging:
\begin{equation}
	\mathbf{912} \to 2\times (\mathbf{1}+\mathbf{35}_s +\mathbf{35}_v +\mathbf{35}_c +\mathbf{350}).
\end{equation}
Obviously the same decomposition is valid also for the complexified versions of E$_7$ and SO(8), whose real sections contain also the gauge group SO(4,4).
In fact, we could also gauge SO(4,4) by introducing magnetic fields coupled to the appropriate SL(8,${\mathbb R}$) generators via a second tensor $\xi$ in the $\mathbf{36}$ of SL(8,${\mathbb R}$),  so that \cite{DallAgata:2011aa}
\begin{equation}
	\label{xicoupling}
	\Theta^{AB\,C}{}_D \propto \delta^{[A}_D \xi_{\vphantom{D}}^{B]C},
\end{equation}
and 
\begin{equation} 
	\label{xitheta}
	\xi = c\, \theta^{-1}
\end{equation} 
in order to satisfy the quadratic constraint
\begin{equation}
	\label{quadconstraint}
	\Theta_M{}^\alpha \Theta_N{}^\beta \Omega^{MN} = 0,
\end{equation}
which is required by consistency of the gauging.
We will refer to the models of eqs.~(\ref{thetacoupling}), (\ref{theta}), (\ref{xicoupling}) and (\ref{xitheta}) as the SO(4,4)$_c$ gaugings.
These are a one-parameter family of gauged supergravity theories with SO(4,4) gauge group, whose details depend on the value of the parameter $c$.
By following the same procedure as in \cite{Dall'Agata:2012bb}, we find that $c$ describes inequivalent theories in the range $c \in [0,\sqrt2-1[$.
Actually, in this case we will argue that the range of inequivalent gaugings is larger than this.
As we will see the scalar potential will change in the full range $c \in [0,1]$, meaning that beyond $c = \sqrt2 -1$ there should be another invariant tensor that allows us to refine our analysis and further distinguish inequivalent models.

As explained in \cite{de Wit:2007mt}, the embedding tensor fixes completely the gauging and it fully determines, among the various couplings, the scalar potential, which is going to be at the center of our analysis.
In fact, from the embedding tensor we can construct the structure constants of the gauge group 
\begin{equation}
	X_{MN}{}^P = \Theta_M{}^\alpha [t_{\alpha}]_N{}^P
\end{equation}
and write the scalar potential in terms of them and of the coset representatives $L$ and their combination ${\cal M} = L L^T$:
\begin{equation}\label{scalar potential}
	V(\phi) = \frac{g^2}{672}\left( X_{MN}{}^{R} X_{PQ}{}^{S} {\cal M}^{MP} {\cal M}^{NQ} {\cal M}_{RS}+ 7 \,X_{MN}{}^{Q} X_{PQ}{}^{N} {\cal M}^{MP} \right).
\end{equation}

Obviously, the generic potential obtained in this fashion is extremely complicated and depends on all 70 scalar fields.
For this reason, we focus on a truncation that can be analyzed more easily.
As in \cite{HullWarner,Dall'Agata:2012bb}, we keep only the scalars that are singlets with respect to a group $G \subset$ SU(8) $\cap$ SO(4,4).
In order to have a limited, but significative number of fields, we first reduced our analysis to the scalar fields that are singlets with respect to an SO(3) $\times$ SO(3) subgroup of SO(4,4) and then further truncate the model by imposing a ${\mathbb Z}_2$ symmetry.
The SO(3) $\times$ SO(3) group is taken by selecting the SO(3) factors coming from the diagonal combination of the two SU(2) factors in the SO(4) $\simeq $ SU(2) $\times $ SU(2) subgroups of SO(4,4).
This leaves 6 scalar fields in total, 2 of which are also invariant under the full compact subgroup SO(4) $\times $ SO(4).
In detail, the 70 scalar fields form the $\mathbf{35}_v$ and the $\mathbf{35}_c$ of SO(8) and in the decomposition SO(8) $\to $ SO(4) $\times$ SO(4) $\to $ SO(3)$\times$SO(3) 
\begin{eqnarray}
	\label{irreps1}
	\mathbf{35}_v &\to& 4 \, (\mathbf{1},\mathbf{1}) + 2 \, (\mathbf{3},\mathbf{1}) + 2 (\mathbf{1},\mathbf{3}) + (\mathbf{3},\mathbf{3}) + (\mathbf{5},\mathbf{1}) + (\mathbf{1},\mathbf{5}), \\[2mm]
		\label{irreps2}
	\mathbf{35}_c &\to& 2 \, (\mathbf{1},\mathbf{1}) +  (\mathbf{3},\mathbf{1}) + (\mathbf{1},\mathbf{3}) + 3\, (\mathbf{3},\mathbf{3}). 
\end{eqnarray}
Looking at the E$_{7(7)}$ generators in the SL(8,${\mathbb R}$) basis, we can see that each of the scalars corresponds to a non-compact generator, which breaks the SO(4,4) gauge group to one of its subgroups as follows:
\begin{equation}\label{singlets}
	\begin{array}{rcll}
	g_1 &=& t^{1234}+t^{5678}, &  {\rm SO}(4)\times {\rm SO}(4); \\[2mm]
	g_2 &=& t_1{}^1+t_2{}^2+t_3{}^3+t_4{}^4-t_5{}^5-t_6{}^6-t_7{}^7-t_8{}^8, & {\rm SO}(4)\times {\rm SO}(4); \\[2mm]
	g_3 &=& t_1{}^1+t_2{}^2+t_3{}^3-t_4{}^4-t_5{}^5-t_6{}^6-t_7{}^7+t_8{}^8, &{\rm SO}(3,1) \times {\rm SO(1,3)};                      \\[2mm]
	g_4 &=& t_4{}^8+t_8{}^4, &{\rm SO}(3,3);                      \\[2mm]
	g_5 &=& t_1{}^1+t_2{}^2+t_3{}^3+t_5{}^5+t_6{}^6+t_7{}^7-3(t_4{}^4+t_8{}^8),  & {\rm SO}(3,3)\times {\rm SO}(1,1); \\[2mm]
	g_6 &=& t^{1238}+t^{4567}, &  {\rm SO}(3,1)\times {\rm SO}(3,1).      
	\end{array}
\end{equation}
It is also clear from these equations that the first and the last generators are in the $\mathbf{35}_c$, while the remaining 4 are in the $\mathbf{35}_v$.
We also see that the only common subgroup preserved by turning on generic expectation values of these scalar fields is SO(3) $\times $ SO(3).

Some of the automorphisms of SO(4,4) are symmetries of the scalar potential. 
Hence, we can perform a further truncation with respect to some discrete $\mathbb Z_2\subset \text{Aut}(\text{SO}(4,4))\cap\text{SL}(8,\mathbb R)$. 
We focus on the ${\mathbb Z}_2$ projection that preserves the $g_5$ and $g_6$ generators, which reveals some interesting new features.
This projection is defined by the element 
\begin{equation}
	Z = \sigma_1 \otimes ({\mathbb 1}_3 \oplus -1),
\end{equation}
acting on the SL(8,${\mathbb R}$) indices as the permutation
\begin{equation}
	\left(\begin{array}{cccccccc}
	1 & 2 & 3 & 4 & 5 & 6 & 7 & 8 \\
	5 & 6 & 7 & -8 & 1 & 2 & 3 & -4
	\end{array}\right).
\end{equation}
This truncation preserves the generators $g_4$, $g_5$ and $g_6$, although only the scalar fields corresponding to $g_5$ and $g_6$ will appear in the potential, because $g_4$ is one of the generators of the SO(4,4) gauge group under which the scalar potential is invariant\footnote{Actually, we could also introduce a further projection by $Z^\prime = \sigma_3 \otimes ({\mathbb 1}_3 \oplus -1)$, which, together with $Z$ generate the discrete group $D_4$ and further restrict the invariant generators to $g_5$ and $g_6$, but this is inessential to our purpose.}.
We will now present the details of the scalar potential and its critical points in this sector.


\section{Analysis of the extrema} 
\label{sec:analysis_of_the_extrema}

\begin{figure*}[t]
	\begin{center}
 \includegraphics[scale=.7]{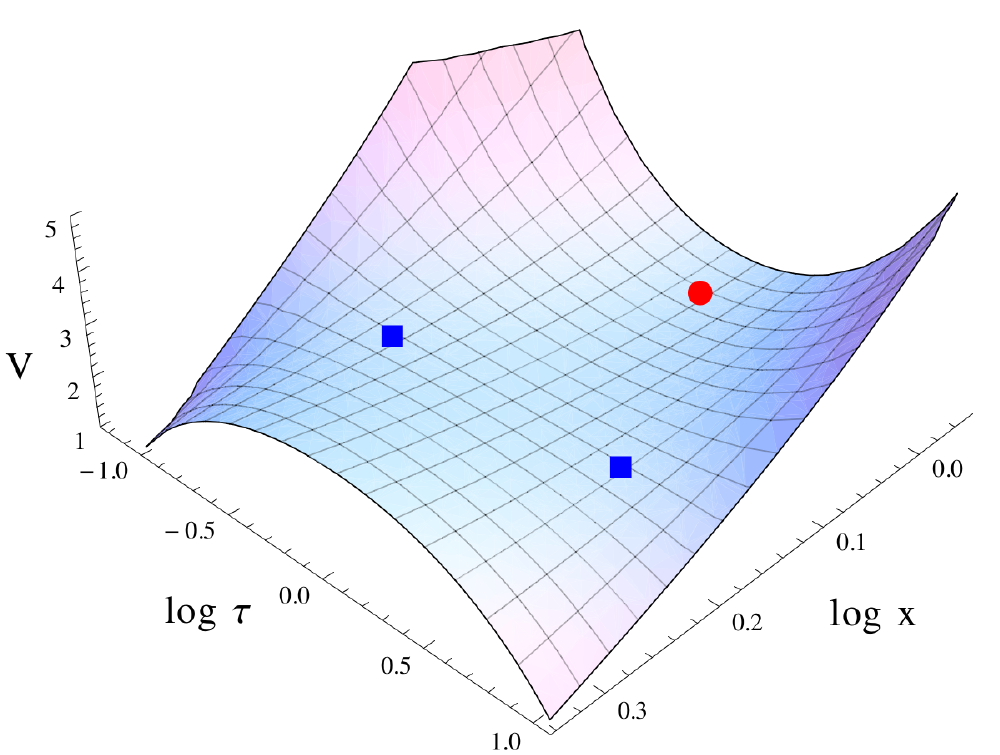} \hspace{10mm}
 \includegraphics[scale=.7]{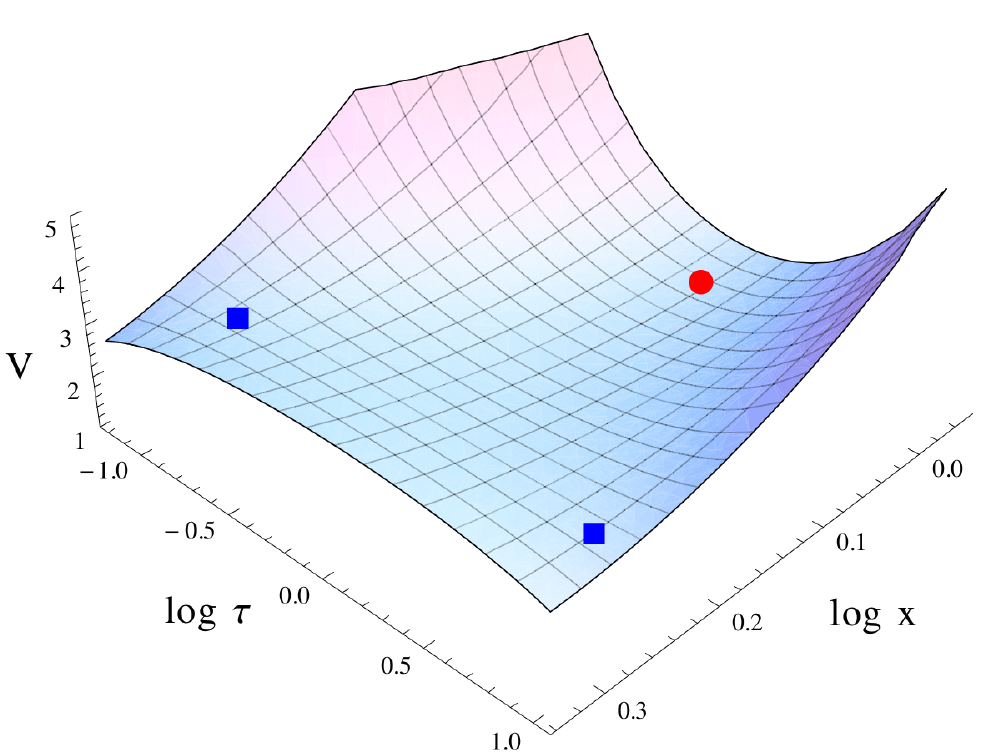} 
 	\end{center}
\caption{\label{fig:potentials} Scalar potential for the SO(4,4)$_c$ model with $c=0$ and with $c=1/3$. The instabilities of the de Sitter vacua represented by blue squares become milder as $c$ approaches the boundary value $\sqrt2-1$. Note that the vacuum represented by a red dot is locally stable in the $\tau$ and $x$ directions.}
\end{figure*}

We now restrict our analysis of the scalar potential (\ref{scalar potential}) to the scalars related to the $g_5$ and $g_6$ generators.
By taking $g_5$ and $g_6$ normalized so that Tr($g_i g_i$) =1, we define the associated coset representative as
\begin{equation}
	L(x, \tau) = \exp\left(\frac{3}{\sqrt2}\,g_5\, \log x +\sqrt{6} \,g_6\, \log \tau\right).
\end{equation}
Obviously, in this parameterization the allowed moduli space is spanned by $x>0$ and $\tau>0$.
We then see that the explicit form of (\ref{scalar potential}) becomes
\begin{equation}
	\begin{array}{rcl}
	V(x, \tau) &=& \displaystyle \frac1{8 \,\tau\, x^{3/2}} \left[c^2 (\tau +1)^2 x^3+3 x^2 \left(2 c^2 (\tau -1)^2-(\tau -6) \tau
   -1\right) \right.\\ [4mm]
   &&\left.-3 x \left(c^2 ((\tau -6) \tau +1)-2 (\tau -1)^2\right)+(\tau
   +1)^2\right].
   	\end{array}
\end{equation}
As depicted in Fig.~\ref{fig:potentials} for two representative choices of $c$, this potential has 3 extrema in the range $c \in [0,\sqrt2 -1[$, all of which have a positive cosmological constant. 

The first critical point is at
\begin{equation}
	x_1 = 1, \qquad \tau_1 = 1.
\end{equation}
At this critical point the potential is\footnote{If one chooses to parameterize $\theta$ and $\xi$ using $\sin \omega$ and $\cos \omega$, as in \cite{Dall'Agata:2012bb}, the value of $V_1$ becomes independent on $\omega$. Recall that one has always the freedom to rescale simultaneously $\theta$ and $\xi$ by changing the value of the gauge coupling constant $g$.}
\begin{equation}
	V_1 = 2 (1+ c^2) >0
\end{equation}
and the gauge group is broken to the SO(4) $\times $ SO(4) subgroup.
We computed also the masses of all the scalar fields (normalized by the cosmological constant), finding the spectrum reported in Table \ref{tab:spectrumc0}.

\begin{table}[h]\renewcommand{\arraystretch}{1.3}\addtolength{\tabcolsep}{-1pt}%
\begin{center}
\begin{tabular}{|c|c|l|}
\hline
	residual G$_{g}$		& $\Lambda$	& $m^2{}_{\text{ (multiplicity)}}$ \\
\hline\hline
	SO(4)$\times$SO(4) & $2(1+c^2)$
	& $0_{(16)},\ 1{}_{(16)},\ 2{}_{(36)},\ -2{}_{(2)} $\\
\hline
\end{tabular}
\end{center}
\caption{Values of the effective cosmological constant and scalar masses in units of the cosmological constant for the critical point at $x=\tau=1$.}
\label{tab:spectrumc0}
\end{table}

The 16 massless fields are all Goldstone bosons for the broken gauge symmetries.
The critical point, however, is not stable because of the 2 tachyonic directions, which have a mass of the order of the cosmological constant (as already found in \cite{HullWarner,Kallosh:2001gr} for $c=0$ and in \cite{Kodama:2012hu} for generic $c$).
It is also interesting to see that, although the value of the cosmological constant changes in a finite range as $c$ varies in the interval $[0,\sqrt2 -1[$, the normalized mass spectrum remains fixed.
This means that these vacua fall in the same class of those in  \cite{Dall'Agata:2012bb,Borghese:2012qm,Kodama:2012hu}, whose mass pattern was explained by the fact that the masses are related to the structure constants of the residual gauge group \cite{Kodama:2012hu}.

The other two critical points are related by parity mapping $\tau \to \frac{1}{\tau}$ (or $\phi \to - \phi$ if we parameterize $\tau = e^\phi$).
They appear at\footnote{For $c=0$ these vacua were also found by T.~Fischbacher, analizing an $N=1$ truncation of this model \cite{private}.}
\begin{equation}
	\label{taucrit}
	\tau_{2,3} = \frac{1}{(1+x)(c^2 x -1)} \left[1-(3+c^2(x-3))x \pm 2 \sqrt2 \sqrt{(c^2-1)x(1-x)(1+c^2 x)}\right]
\end{equation}
and at the real positive root $x= x_*$ of the equation
\begin{equation}
	1 - 3 (c^2 -2) x + 3(2c^2 -1) x^2 + c^2 x^3 = 0,
\end{equation}
which gives $\tau_{2,3}>0$ when inserted in (\ref{taucrit}).
Note that for $c \to \sqrt2 -1$ the position of the vacua in the $\tau$ coordinate approaches the boundary of the moduli space ($\tau \to 0$ and $\tau \to \infty$ for the two vacua, respectively) and for $c \geq \sqrt2 -1$ we are left only with the central vacuum.
Hence we constrain our analysis to the interval $c \in [0,\sqrt2 -1[$.
For $c \in [\sqrt2-1,1]$ we still have legitimate models, but the scalar potential has only the vacuum at the center of the moduli space ($x=\tau=1$).
In order to produce compact expressions for the various quantities we are going to compute in the following, it is useful to express everything in terms of $x_*$.
This value ranges between $x_*=1+\frac{2}{\sqrt3}$ at $c=0$ and $x_* \to 3+ 2 \sqrt2$ for $c \to \sqrt2 -1$, and one can recover the value of $c$ that specifies the SO$(4,4)_c$ model from:
\begin{equation}
 c = \frac{\sqrt{3x_*^2-6x_*-1}}{\sqrt{x_*}\sqrt{x_*^2+6x_*-3}}.
\end{equation}
The critical values of $\tau$ then become
\begin{equation}
	\tau_{2,3} = \frac{-3\pm 2\sqrt2 + 2 x_* - (3 \pm 2 \sqrt2) x_*^2}{x_*^2-6x_*+1}.
\end{equation}

The value of the cosmological constant of these new vacua is once more positive and depends on the deformation parameter $c$.
The scalar potential at the new critical points is
\begin{equation}
	V(x_*,\tau_{2,3}) = 3 \frac{(x_*-1)(x_*+1)^3}{x_*^{3/2}(x_*^2+6x_*-3)}
\end{equation}
and varies between 
\begin{equation}
	V_{c=0} = 2 \sqrt{6 \sqrt3-9} \leq V(x_*,\tau_{2,3}) < 12(\sqrt2-1) = V_{c\to\sqrt2-1}.
\end{equation}
We stress that since $\tau$ is associated to $g_6\, \slashed{\in}\, {\mathfrak sl}(8,{\mathbb R})$ and these vacua appear at $\tau \neq 1$, they could not have been found in the analyses of \cite{DallAgata:2011aa,Kodama:2012hu}, which considered only points connected to the origin of the moduli space by SL$(8,{\mathbb R})$ transformations.

\begin{figure*}[h]
	\begin{center}
 \includegraphics[scale=.27]{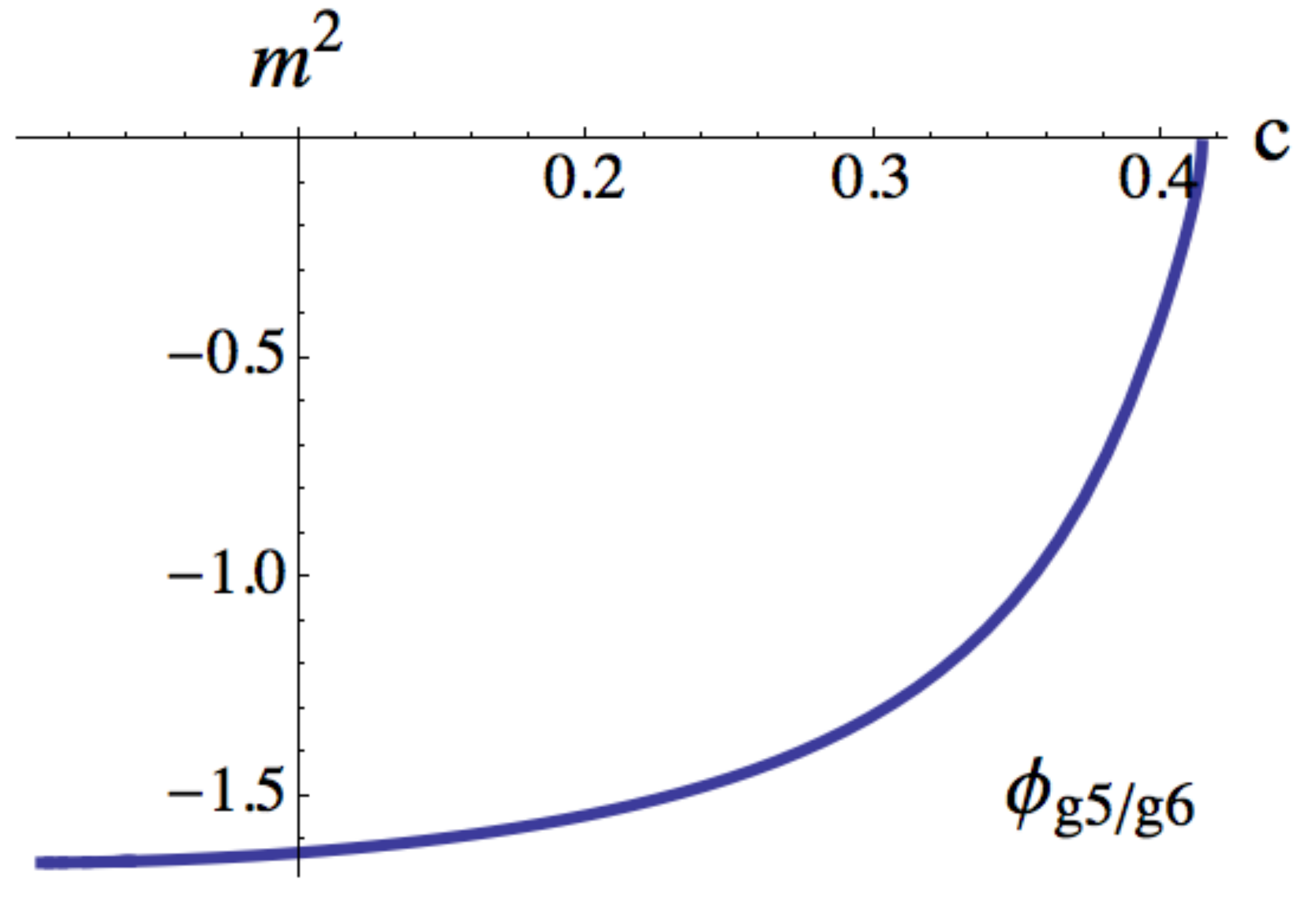} 
 \includegraphics[scale=.27]{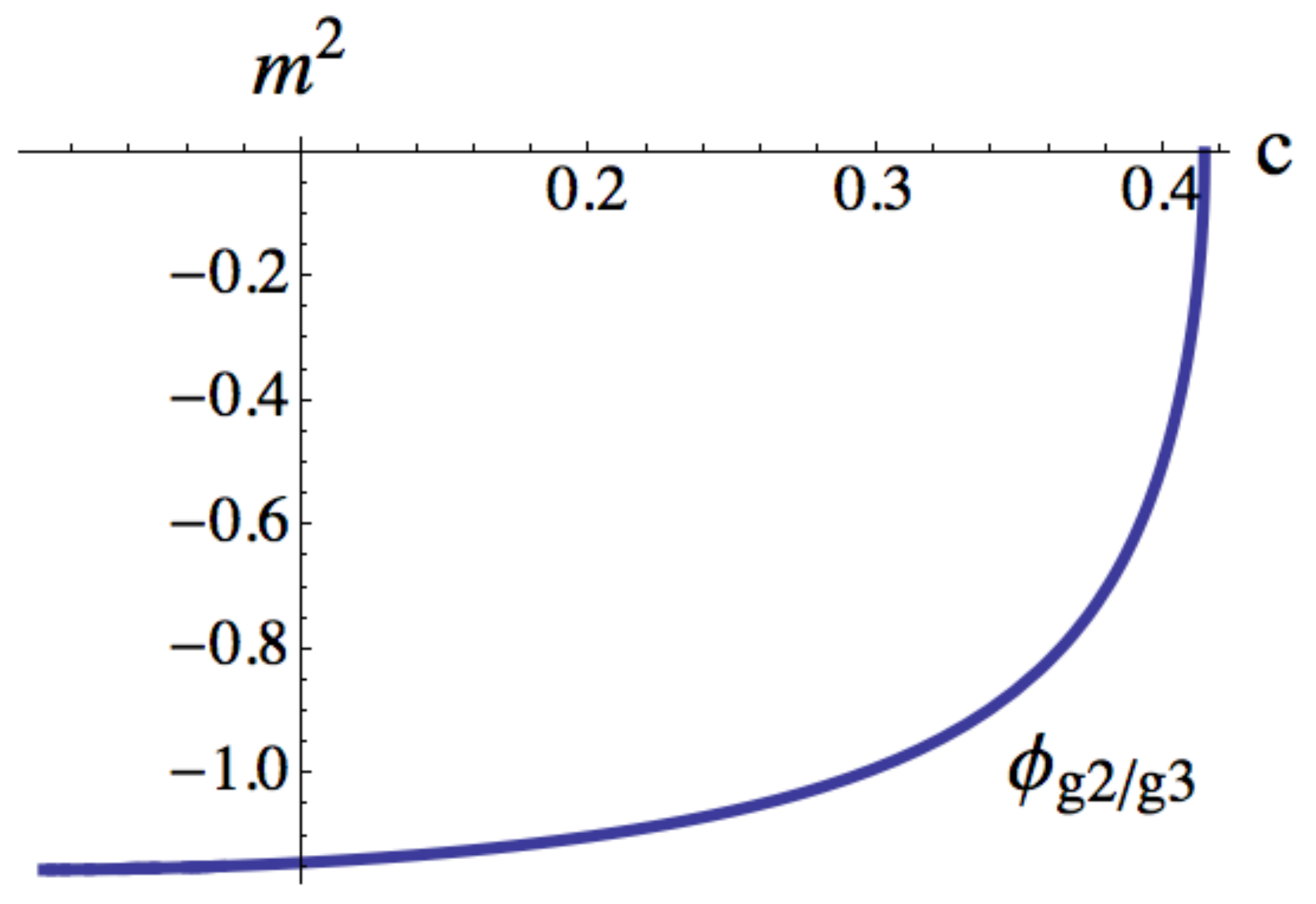} 
 \includegraphics[scale=.27]{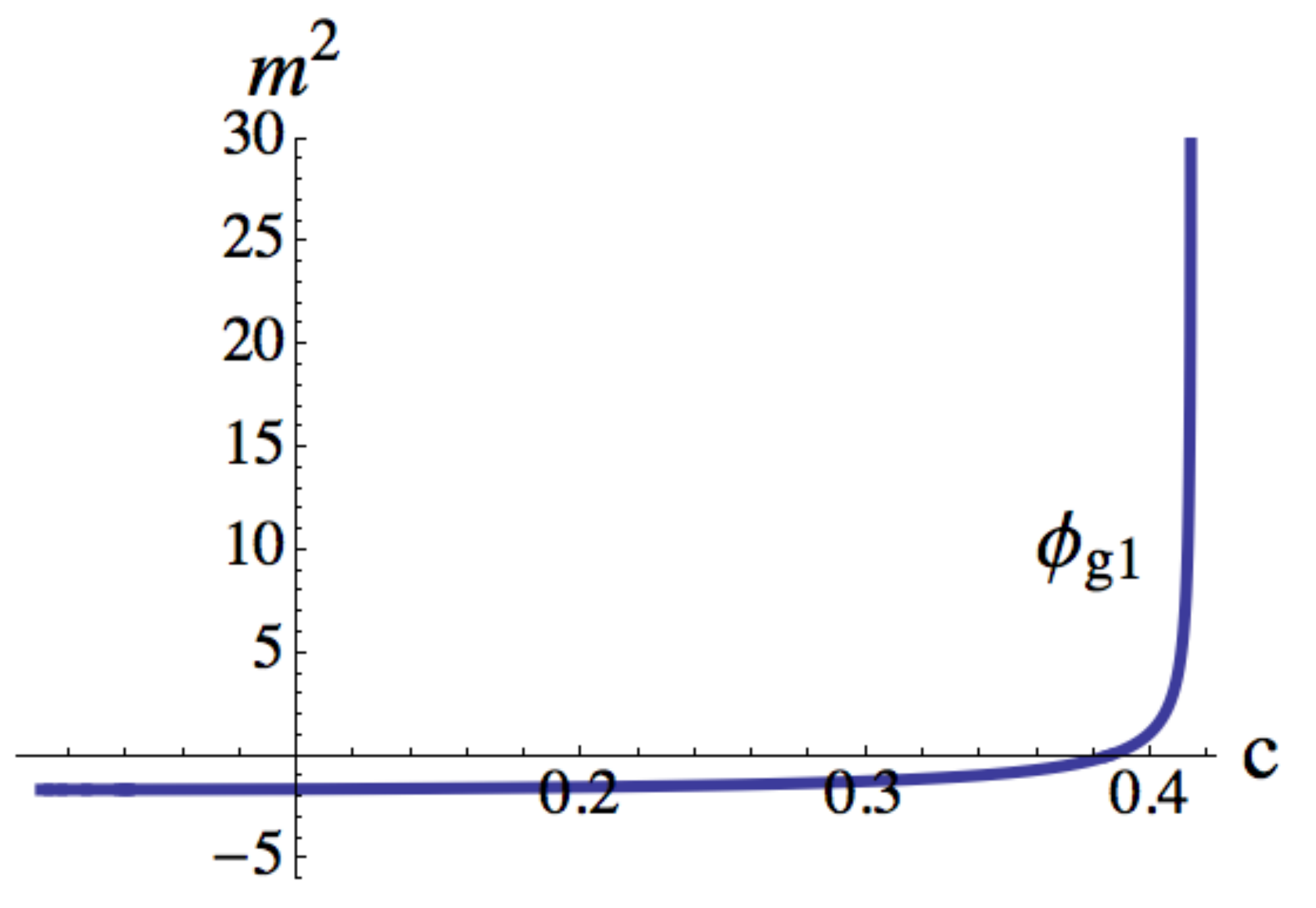}	\end{center}
\caption{\label{fig:3masse} Value of the squared masses normalized in units of the cosmological constant in the range $0\leq c <\sqrt2 -1$ for the three tachyonic directions at $c=0$. The first two approach zero as $c$ approaches $\sqrt2 -1$. The third one starts negative at $c=0$, becomes positive and then explodes as $c \to \sqrt2 -1$.}
\end{figure*}

Also for these vacua we can compute the full mass spectrum analytically.
We always have 6 massless vectors, which implies that we also find 22 massless scalar fields corresponding to Goldstone bosons of the broken gauge symmetry, which is now reduced to SO(3) $\times $ SO(3).
All of the other scalar squared masses are always positive except for three of them, which are associated to SO(3) $\times$ SO(3) singlets, which are specific combinations of those in (\ref{singlets}).
One of them corresponds to the direction specified by the generator $g_1$ and is tachyonic only for $x_* < 2+\sqrt3$, while it blows up as we approach the boundary, i.e.~$c \to \sqrt2 -1$:
\begin{equation}
	m^2_{\phi_{g_1}} = - 4 \, \frac{x_*^2 - 4 x_* +1}{x_*^2-6 x_* +1}.
\end{equation}
The other two tachyonic fields maintain a negative mass squared over the whole allowed range for $c$ and correspond to directions that are mixtures of the generators $g_5$, $g_6$, for which
\begin{equation}
	m^2_{\phi_{g_{5/6}}} =-\frac23\, \frac{3 - 10 x_* + 3 x_*^2 -2 \sqrt{3-24 x_* + 58 x_*^2 - 24 x_*^3+ 3 x_*^4}}{x_*^2-6 x_* +1},
\end{equation}
and of $g_2$, $g_3$, for which
\begin{equation}
	m^2_{\phi_{g_{2/3}}} = -\frac13\,\frac{3 - 2 x_* + 3 x_*^2 - \sqrt{33-300 x_* + 934 x_*^2 - 300 x_*^3+ 33 x_*^4}}{x_*^2-6 x_* +1}.
\end{equation}
As $c$ approaches the boundary, the combinations defining the tachyons are directed towards $g_6$ and $g_3$, which means that the $\tau$ field captures the most dangerous direction in the potential.
However, the interesting point for our analysis is that when $c \to \sqrt2 -1$ the value of the masses of these two tachyons tends to zero.
We display the value of their normalized mass as a function of $c$ in Fig.~\ref{fig:3masse}.
Since the value of the cosmological constant stays finite in the same region of parameters, we can see that by choosing $c$ close enough to its boundary value we have vacua where most of the directions are stable and where the tachyonic directions have the slow-roll parameter $\eta$, determined by the ratio of the physical tachyonic masses and the cosmological constant, as small as we like.
Although we did not perform an extensive search for other critical points of the scalar potential in the full 70-dimensional scalar field space, the potential is unbounded from below and, generically, we expect that the flow along the $\tau$ direction leads to infinitely negative values of the potential itself.

In order to see the degeneracies of the various scalar masses, we report in Table \ref{tab:otherdS} the spectrum at $c=0$ as well as the limiting values of the cosmological constant and of the normalized masses for $c \to \sqrt2 -1$.
As expected such a degeneracy follows the representation pattern given in Eqs.~(\ref{irreps1})--(\ref{irreps2}).
The Goldstone bosons can be identified as the scalar fields in the same representations as the vector fields acquiring masses in the process of gauge symmetry breaking.
The 28 vector fields are in the representations determined by the decomposition:
\begin{equation}
	\mathbf{28} \to 3 \times \left[(\mathbf{3},\mathbf{1}) + (\mathbf{1},\mathbf{3})\right] + (\mathbf{1},\mathbf{1}) + (\mathbf{3},\mathbf{3}).
\end{equation}
The residual gauge fields of SO(3) $\times$ SO(3) are in the $(\mathbf{3},\mathbf{1}) + (\mathbf{1},\mathbf{3})$ representation and therefore the other 22 are massive and eat the scalar fields in the corresponding representations.
This leaves the remaining fields in representations leading to the degeneracies in Table \ref{tab:otherdS}, with the notable additional degeneracy between a set of scalars in the $(\mathbf{3},\mathbf{3})$ and another in the $(\mathbf{3},\mathbf{1}) + (\mathbf{1},\mathbf{3})$.
 
\begin{table}[h!]\renewcommand{\arraystretch}{1.3}\addtolength{\tabcolsep}{-1pt}%
\begin{center}
\begin{tabular}{|c|c|c|}
\hline
	$c$ value		& $\Lambda$	& $m^2{}_{\text{ (multiplicity)}}$ \\
\hline\hline
	0 & $2 \sqrt{6 \sqrt3-9}$
	& $
	\begin{array}{ccc}
	0_{(22)}, & \frac{2}{\sqrt3}(1+\sqrt3){}_{(10)}, & \frac23(2+\sqrt3){}_{(9)},  \\[2mm]
   \frac12(3+\sqrt3){}_{(15)}, & \frac{4}{\sqrt3}{}_{(1)}, &	\frac23(1+\sqrt3){}_{(9)}, \\[2mm]
  -1 +\frac{1}{\sqrt3}+\sqrt{\frac23(4-\sqrt3)}{}_{(1)},&  -\frac{2}{\sqrt3}{}_{(1)}, & 	\frac12(-5+\sqrt3){}_{(1)}, \\[2mm]
  -1 +\frac{1}{\sqrt3}-\sqrt{\frac23(4-\sqrt3)}{}_{(1)}& &
	\end{array}$  \\[2mm]
\hline\hline
	$\sqrt2 -1$ & $12(\sqrt2-1)$
	& $
	\begin{array}{ccc}
	0_{(24)}, & {1}_{(9)}& +\infty{}_{(37)}
	\end{array}$  \\[2mm]
\hline
\end{tabular}
\end{center}
\caption{Values of the effective cosmological constant and scalar masses in units of the cosmological constant for the critical points ($x_*$, $\tau_{2,3}$) at two different values of $c$.
The last row should be interpreted as the limiting value of the cosmological constant and of the normalized masses as $c \to \sqrt2 -1$.}
\label{tab:otherdS}
\end{table}


\section{Comments and Conclusions} 
\label{sec:comments}

Summarizing, we presented a simple SO(4,4)$_c$ gauged supergravity model that allows for unstable de Sitter vacua with arbitrarily small slow-roll parameter $\eta$.
This is the first instance of vacua of this type in maximal supergravity and provides a counterexample to the intuition built so far on the existing vacua, which all had unstable scalar fields with tachyonic masses of the order of the cosmological constant.

This makes even more compelling a more general analysis, which could provide a no-go theorem for meta-stable de Sitter vacua, or finally provide examples of meta-stable vacua in the maximally symmetric theory.
We are confident that the new parameter-deformed theories of \cite{Dall'Agata:2012bb} will provide a good environment to look for such vacua.
In fact, the example we provided is also the first one where not only the value of the cosmological constant and the positions of the vacua change when introducing the deformation parameter, but also the masses of the scalar fields.
We plan to revisit the models that include known unstable de Sitter vacua to see if the introduction of this parameter makes some of them metastable.

As we saw in the previous section, the new de Sitter vacua appear at different values of $\tau$, approaching the boundary of the moduli space as $c \to \sqrt2 -1$.
It is actually interesting to see that, following the same approach as in \cite{DallAgata:2011aa,inprogress}, we can obtain a contraction of the original model that displays a Minkowski vacuum if we take the $\tau \to 0$, $c \to \sqrt2 -1$ limit at the same time as performing a rescaling of the gauge coupling constant by $g \to \tau g$.
The resulting theory has a gauge group that is a contraction of the original one, namely SO(3,1) $\times $ SO(1,3) $\ltimes$ $T^{16}$, and its scalar potential contains a critical point with vanishing cosmological constant.
The vacuum fully breaks supersymmetry and the gauge group is also broken to SO(3) $\times $ SO(3).
This model provides the first example of a Minkowski vacuum with a residual gauge group that does not have abelian factors.

It is obviously interesting to explore the full moduli space of this model and to investigate its relation with the recent analogous vacua studied in \cite{Dall'Agata:2012cp}, and we plan to report on this in a forthcoming publication.



\bigskip
\section*{Acknowledgments}

\noindent We would like to thank R.~Kallosh, M.~Trigiante and especially F.~Zwirner for interesting discussions and T.~Fischbacher for useful correspondence.
G.D. would also like to thank H.~Kodama, M.~Nozawa and all participants of the workshop ExDiP2012 “Superstring Cosmophysics” for valuable discussions. 
This work is supported in part by the ERC Advanced Grant no. 226455, \textit{``Supersymmetry, Quantum Gravity and Gauge Fields''} (\textit{SUPERFIELDS}), by the European Programme UNILHC (contract PITN-GA-2009-237920), by the Padova University Project CPDA105015/10, by the MIUR-PRIN contract 2009-KHZKRX and by the MIUR-FIRB grant RBFR10QS5J.


\end{document}